\documentclass[12pt]{iopart}

\usepackage{cite}
\usepackage{graphicx}
\usepackage{bm}
\usepackage{upgreek}
\usepackage{soul}

\soulregister\cite7
\soulregister\ref7

\begin{document}

\title[Wang X et. al.]{Ultra-high harmonic conversion of a seeded free-electron laser via harmonic optical klystron}

\author{Xiaofan Wang$^{1 \ast}$, Li Zeng$^{1}$, Weiqing Zhang$^{2 \ast}$, Xueming Yang$^{1,2,3}$}

\address{$^1$ Institute of Advanced Science Facilities, Shenzhen 518107, China}
\address{$^2$ Dalian Institute of Chemical Physics, Chinese Academy of Sciences, Dalian 116023, China}
\address{$^3$ Southern University of Science and Technology, Shenzhen 518055, China.}

\ead{wangxf@mail.iasf.ac.cn, weiqingzhang@dicp.ac.cn}
\vspace{10pt}

\begin{indented}
\item[]May 2023
\end{indented}

\begin{abstract}

External seeded free-electron lasers (FELs) are compelling tools for generating fully coherent EUV and soft X-ray radiations. Echo-enabled harmonic generation (EEHG), the most typical representative of external seeded FELs, has witnessed a remarkable growth of fully coherent FELs in the last decade, continuously evolving towards higher harmonic conversions and shorter wavelengths. Ultra-high harmonic generation is imperative in the field of FELs. This paper presents a novel method for generating FEL radiation with ultra-high harmonic conversion, utilizing harmonic optical klystron in combination with EEHG. This method can effectively increase the harmonic conversion order to about 90. Theoretical analysis and numerical simulations show that intense and almost fully coherent FEL pulses can be generated with a wavelength of 3 nm. At the same time, the seed laser intensity required by this scheme is lower compared to nominal EEHG, thus facilitating the generation of high-repetition-rate seeded FELs.

\end{abstract}
\maketitle
%
%
%
%
%

\section{Introduction}

X-ray free-electron lasers (FELs) worldwide \cite{ackermann2007operation,emma2010first,ishikawa2012compact,Allaria:ig5031,kang2017hard,decking2020mhz,prat2020compact} have opened up new frontiers of ultra-fast and ultra-small sciences on molecular and atomic length scales \cite{RevModPhys.88.015007,HUANG2021100097}. Most of the operational FEL facilities are based on the mode of self-amplified spontaneous emission (SASE)\cite{kondratenko1980generation,bonifacio1984collective}, where lasing happens when an electron beam approaching the speed of light passes through a magnetic structure (undulator). Subject to the inherent beam noise, SASE FELs have chaotic temporal profiles and noisy spectra. Huge efforts have been made to develop fully coherent FELs. The self-seeding scheme \cite{feldhaus1997possible,amann2012demonstration,ratner2015experimental}, which adopts an optical monochromator during the FEL amplification process, purifies the radiation spectra but reserves the SASE power fluctuations.

Seeded FELs, which utilize the technique of optical-scale beam manipulation with external ultraviolet lasers, are alternative ways to improve temporal coherence and suppress fluctuations of SASE FELs \cite{yu1991generation,yu2000high,penco2020enhanced,YU199796,PhysRevSTAB.16.020704,allaria2013two,stupakov2009using,xiang2009echo,PhysRevLett.115.114801,PhysRevAccelBeams.23.060701,PhysRevLett.105.114801,PhysRevLett.108.024802,zhao2012first,PhysRevSTAB.17.070702,hemsing2016echo,feng2019coherent,ribivc2019coherent,Feng:22,feng2010hard,ZHAO2016720,FAN2022166241,photonics8020044}. The inheritance of properties between the radiation and the seed laser makes the seed FELs the only devices currently capable of generating laser-like X-ray pulses with Fourier transform limits, stability, repeatability and controllability. Based on these characteristics, seeded FELs are of great value in many research fields, e.g. spectroscopy \cite{petrillo2019high} and coherent diffraction imaging \cite{pedersoli2013mesoscale,capotondi2015multipurpose,helfenstein2017coherent}.

One classical approach of seeded FELs is the high-gain harmonic generation (HGHG) \cite{yu1991generation,yu2000high}. HGHG consists of two undulators and one dispersive section. In the first undulator (modulator), an external seed is introduced to interact with the electrons, generating a sinusoidal energy modulation of the e-beam at the optical wavelength scale. The dispersion element (chicane) is followed up to transform the energy modulation into an associated spatial density modulation. Then, the micro-bunched electron beam traverses the second undulator (radiator), which is tuned to a high harmonic of the seed laser, to generate coherent radiation at a short wavelength. HGHG FELs have high temporal coherence while low conversion efficiency at frequency. The wavelength coverage is limited to the extreme ultraviolet (EUV) range~\cite{penco2020enhanced}.  Multiple stages of HGHG with the “fresh bunch” technique are generally required~\cite{YU199796,PhysRevSTAB.16.020704,allaria2013two} to reach the X-ray regime.

Later, echo-enabled harmonic generation (EEHG)~\cite{stupakov2009using,xiang2009echo} was proposed to drive seeded FELs to shorter wavelengths in a single-stage setup. Two modulators and two chicanes are employed to realize nonlinear phase space manipulation, resulting in unparalleled up-conversion efficiency and enhanced robustness of the output characteristics to electron beam imperfections \cite{PhysRevLett.115.114801,PhysRevAccelBeams.23.060701}.
 Huge efforts are being made worldwide for the demonstration of EEHG to generate fully coherent laser pulses \cite{PhysRevLett.105.114801,PhysRevLett.108.024802,zhao2012first,PhysRevSTAB.17.070702,hemsing2016echo,feng2019coherent,ribivc2019coherent}. At FERMI, intense and almost fully coherent FEL pulses were produced at a wavelength of 5.9 nm. Coherent emission at harmonics in the range from 84 to 101 was also observed, indicating the potential of EEHG FELs to cover X-ray spectral region \cite{ribivc2019coherent}.

The cascade of EEHG with HGHG, termed as echo-enabled harmonic cascade (EEHC) \cite{Feng:22}, holds the capability of further increasing the up-frequency conversion efficiency. EEHC combines the difficulties of EEHG and cascade HGHG. It uses EEHG as a seed to drive HGHG in the second stage. EEHC has been adopted as one of the leading operation modes for the SXFEL~\cite{app7060607} and SHINE~\cite{zhao2018xfel}. The principle of proof experiment in SXFEL demonstrated the feasibility of EEHC to generate intense and nearly transform-limited soft X-ray radiation with a wavelength of 8.8 nm.
 The use of EEHC to generate shorter-wavelength radiation pulses still requires further experimental confirmation. Multiple stages of EEHG, termed as echo-enabled staged harmonic generation (EESHG) \cite{feng2010hard,ZHAO2016720,FAN2022166241}, is another cascade option of EEHG. Although theoretically, EESHG has unrivaled up-frequency conversion efficiency, the complex physical design hinders its practical implementation.

Harmonic cascade is another method to increase the harmonic number of a single-stage EEHG \cite{photonics8020044}. This method uses EEHG as the first stage to generate micro-bunching at a certain harmonic. A reverse tapered undulator \cite{PhysRevSTAB.9.050702}, which is resonant at this harmonic, is employed to enhance its fundamental and harmonic bunching while suppressing the energy spread growth. FEL pulses at high harmonic are amplified in the final radiator. The hypothesis of long-term maintenance of the EEHG band structure in the reverse taper undulator, however, awaits future theoretical investigations.

Here, a novel approach is proposed to increase the harmonic number of a seeded FEL by combining the harmonic optical klystron with EEHG. Optical klystron (OK) was initially invented for controlling FEL gain in oscillators \cite{vinokurov1977maximum,dattoli2003mopa}. The basic scheme of OK consists of two modulators and one chicane. The layout is similar to HGHG, with the difference that HGHG has a radiator while OK does not. In the second modulator, OK enhances the electron bunching and the coherent emission. If the resonant wavelength of the second modulator is a harmonic of that in the first one, this scheme is called harmonic optical klystron (HOK). Nowadays, OK is not only used in a SASE FEL to speed up the FEL lasing process \cite{penco2015experimental,photonics4010015}, but also used in seeded FELs to relax the power requirement of the seed laser ~\cite{dunning2011design,paraskaki2021high,yan2021self,10.1117/1.APN.2.3.036004}.

The proposed technique generates ultra-high harmonic radiation by combining HOK with EEHG. This technique first uses HOK to generate micro-bunching at a low harmonic, i.e., the second harmonic of the seed. Then, a single-stage EEHG, which is optimized at the forty-fifth harmonic, is used to perform the overall ninetieth harmonic jump. Simulations demonstrate that the proposed technique could generate soft X-ray radiation at a wavelength of 3 nm, a peak power of several gigawatts, and a pulse duration of about 15 femtoseconds when the wavelength of the seed laser is 270 nm.

\section{Physical principle of the proposed technique}

\begin{figure*}[!htb]
\begin{indented}
\item[]
    \includegraphics[width=1\linewidth]{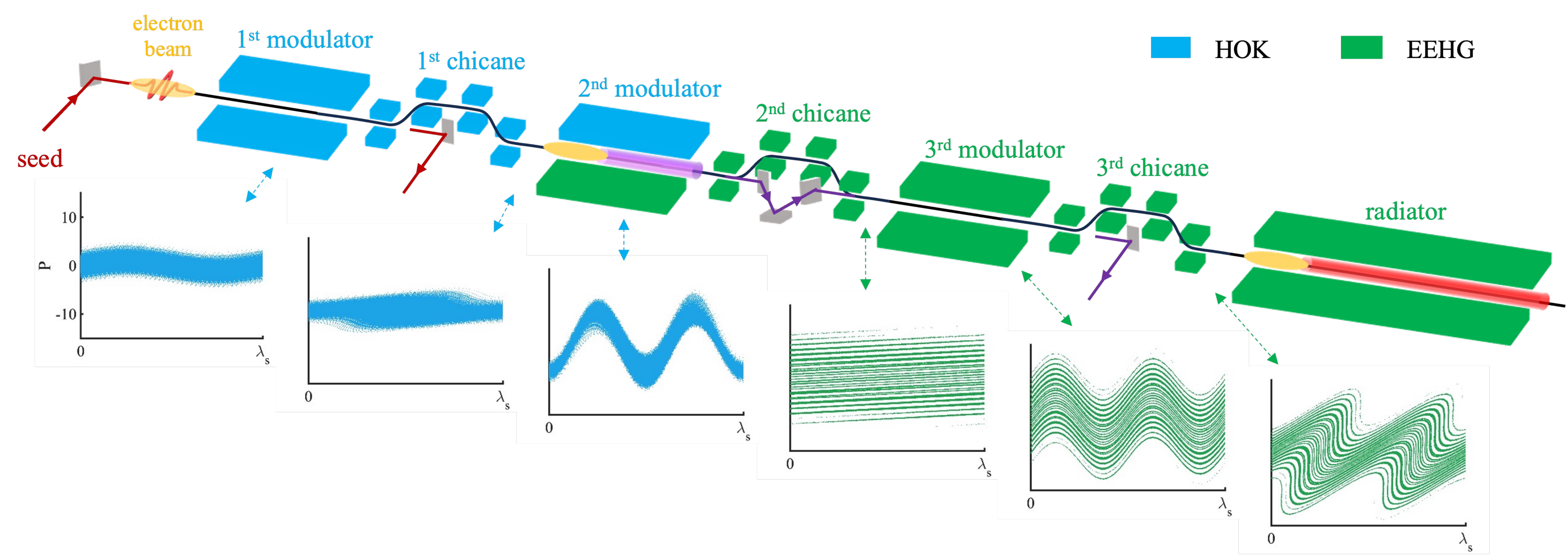}
    \centering
    \caption{Schematic layout and representative longitudinal phase space evolution of HOK-EEHG.}
    \label{fig1}
    \end{indented}
\end{figure*}

For convenience, the proposed technique is named HOK-EEHG in the following content. 
Figure \ref{fig1}
illustrates the schematic layout and representative longitudinal phase space evolution of HOK-EEHG. Unlike the nominal EEHG, HOK-EEHG possesses three modulators and three chicanes but no second seed laser. The first two modulators and the first chicane make up the HOK, while the last two modulators, the last two chicanes, and the radiator make up the EEHG. HOK not only achieves the second harmonic conversion of the external seed but also performs the first-stage energy modulation of EEHG. This is because electron modulation and harmonic emission amplification are achieved simultaneously in the second modulator. Then, the electrons, along with the second harmonic emission, are led to the third modulator for the second-stage energy modulation of EEHG. With optimized dispersion strength of the last two chicanes, the electron beam forms micro-bunching that has frequency components at 90th (90=2*45) harmonic of the seed laser, where 2nd harmonic conversion is done by HOK, while 45th harmonic conversion is done by EEHG. 

The longitudinal phase space evolution in HOK-EEHG is also shown. The horizontal axis represents the longitudinal position of the electron beam, and the vertical axis represents the dimensionless electron beam energy. Since laser-electron manipulation occurs on an optical wavelength scale, only one seed laser wavelength is chosen here as a reference for the horizontal axis. Initially, the electron beam is uniformly distributed along the longitudinal direction. In the first modulator, the seed laser with a wavelength of $\lambda_s$ modulates the electron beam and imprints a weak energy modulation on the e-beam. Then, the chicane modifies the electron beam distribution to maximize the bunching at the second harmonic. The micro-bunched electron beam produces second-harmonic coherent emission with a wavelength of $\lambda_s/2$ in the second modulator with exponential gain, while coherent emission in turn accomplishes spontaneous modulation onto the electron beam. It should be noted that the phase space distribution at the end of the HOK is not perfectly sinusoidal. This effect may affect the case where the follow-up is HGHG, but for the case where EEHG followed-up, the strong dispersion tears the energy into strips and is therefore insensitive to whether the energy modulation is perfectly sinusoidal. In the second strong chicane, electrons of different energies move relative to each other, forming a striated phase space. In the third modulator, the electron beam is again modulated by the coherent radiation from the second modulator. After rotating the phase space in the third chicane, the energy modulation is finally transformed into a periodic density modulation with high frequency components. The micro-bunched electron beam is then directed to the radiator that is tuned to emit FEL pulses at the 90th harmonic of the seed laser.

The advantages of this scheme are: (1) only one seed laser is required; (2) the modulation lasers in EEHG are halved in wavelength by means of HOK, so this scheme can produce coherent radiation at shorter wavelength with the same EEHG harmonic conversion order; (3) halving the wavelengths of the modulation lasers in EEHG also results in halving the dispersion intensities required for energy-to-position conversion, which is good for reducing the coherent synchrotron radiation in the chicane; and (4) two seed lasers used in EEHG are self-generated by the electron beam, which is favorable for the generation of FEL pulses without externally introduced timing jitter.

\section{Start-to-end three-dimensional simulations}

\begin{figure}[!htb]
\begin{indented}
\item[]
    \includegraphics[width=\linewidth]{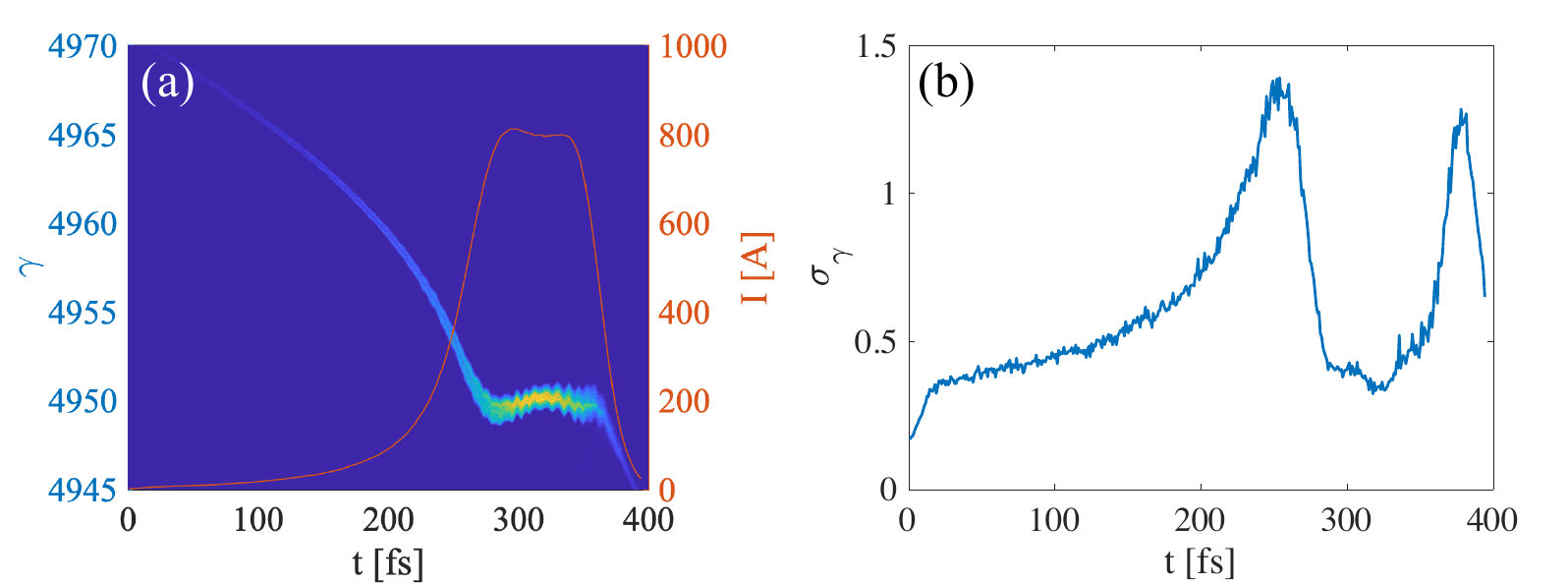}
    \caption{(a) Current profile along with the longitudinal phase space of the e-beam; (b) energy spread along the e-beam.}
    \label{fig2}
    \end{indented}
\end{figure}

The feasibility of the proposed method is demonstrated with start-to-end three-dimensional simulations~\cite{REICHE1999243}. Beam parameters of $\rm{S}^3$FEL \cite{wang14th} are utilized here. $\rm{S}^3$FEL is a newly proposed high repetition rate X-ray FEL facility, with a beam energy of 2.5 GeV, a peak current of 800 A, an energy spread of 0.19 MeV, and normalized emittance of 0.5 mm·mrad. $\rm{S}^3$FEL aims to generate EUV and X-ray FELs between 1 and 30 nm under the working principle of SASE and seeded FELs. The seeding lines are planned to generate intense FEL radiation with the shortest wavelength of about 3 nm. Figure~\ref{fig2} presents the current profile along with the longitudinal phase space and the energy spread distribution.
From the figure, one can get that the electron beam has a time duration of about 110 fs (FWHM). It can be seen from Fig.~\ref{fig2}(b) that slice energy spread in the current plateau region also varies with time, which contributes to the generation of ultrashort FEL pulses.

\begin{figure*}[!htb]
\begin{indented}
\item[]
    \includegraphics[width=1\linewidth]{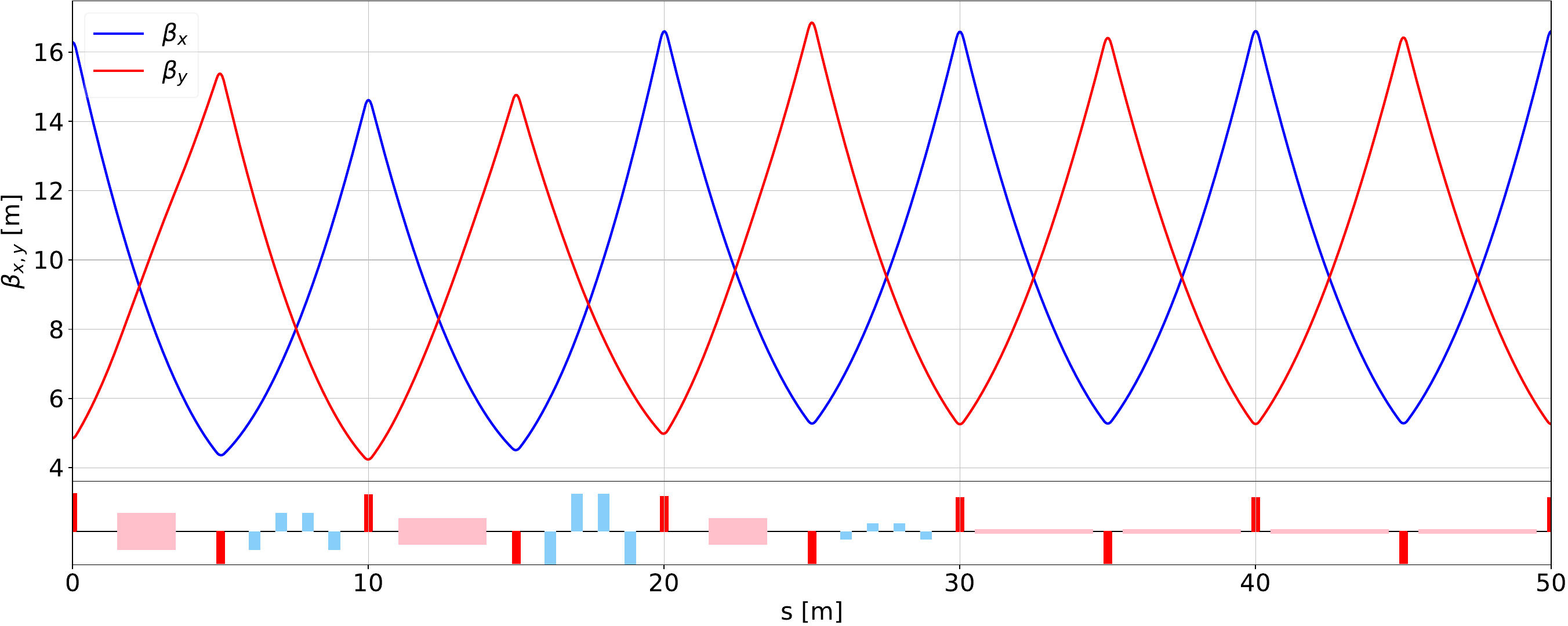}
    \caption{Beamline of HOK-EEHG and $\beta$-functions evolution.}
    \label{fig3}
    \end{indented}
\end{figure*}

The brief layout of the HOK-EEHG beamline, together with the evolution of the $\beta$-functions are plotted by Ocelot \cite{{AGAPOV2014151}} in Fig.~\ref{fig3}. The average value of the $\beta$-function for the whole beamline is around 10, favoring energy modulation and FEL radiation. All the undulators used here are the out-of-vacuum and variable gap type. Three modulators have segment lengths of 2 m, 3 m, and 2 m, respectively. The period lengths are all 9 cm. The first modulator resonants at 270 nm, while the second and third modulators resonate at the second harmonic of the seed laser with a wavelength of 135 nm. The radiator has a resonant wavelength of 3 nm, a period length of 4.3 cm, and a segment length of 4 m. 

\begin{figure}[!htb]
\begin{indented}
\item[]
    \includegraphics[width=\linewidth]{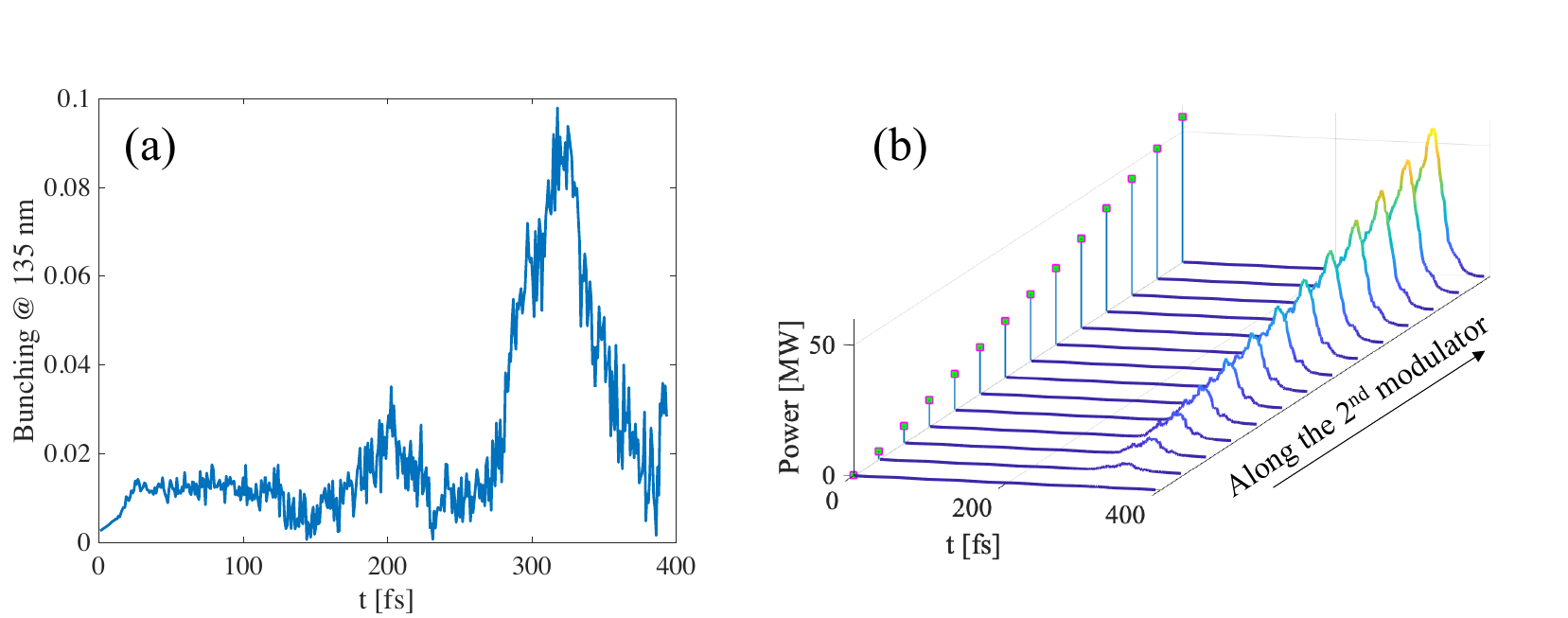}
    \caption{(a) Bunching factor at 135 nm along the beam; (b) evolution of the coherent harmonic radiation power along the second modulator.}
    \label{fig4}
    \end{indented}
\end{figure}

The external seed laser is provided by the third harmonic generation of a Ti:sapphire laser. It has a wavelength of 270 nm, a peak power of 1.78 MW, and a pulse length of 350 fs (FWHM). It is also assumed that this is a Gaussian laser with an optical waist position in the middle of the modulator and a radial rms beam size of 150 $\upmu$m. The spatial variation of the laser field directly leads to the fact that the electrons at the radial edge receive different modulation depths than the electrons in the core region. Considering this phenomenon, the energy band structure required for EEHG will be somewhat affected and the bunching factor will be lower than the theoretical prediction. In the three-dimensional simulations, the transverse and longitudinal variations of the light field, either the external seed laser or the second harmonic emission generated by the e-beam, are thoroughly considered. After the first modulator, the seed laser would induce 0.21 MeV energy modulation on the e-beam, which is about 1.1 times the energy spread. When the momentum compaction of the first chicane is set to 0.36 mm, the energy modulation transfers to spatial density modulation. The maximum bunching factor at 135 nm along the electron beam is about 9\%. Due to the drastic change in energy spread with time, the duration of the beam bunching is 47.7 fs (FWHM), which is shorter than that of the electron beam. Driven by the density modulation, coherent radiation at the second harmonic is emitted from the electron beam in the second modulator. Radiation power grows exponentially and reaches about 55.5 MW at the exit of the modulator. The time duration of the radiation pulse is about 40.8 fs (FWHM). Figure~\ref{fig4} illustrates these results.

\begin{figure}[!htb]
\begin{indented}
\item[]
    \includegraphics[width=\linewidth]{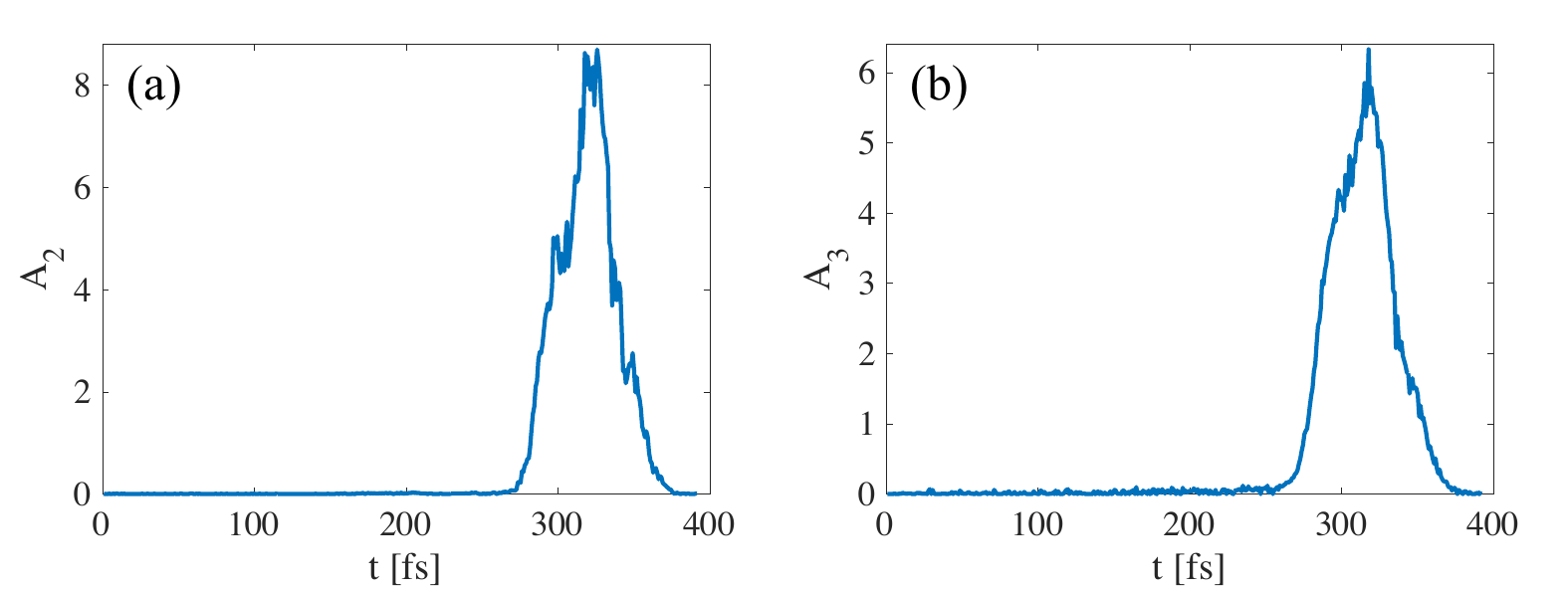}
    \caption{Laser-induced modulation amplitude in the second modulator (a) and the third modulator (b).}
    \label{fig5}
\end{indented}
\end{figure}

Electron modulation and second harmonic amplification are realized simultaneously in the second modulator. Coherent harmonic emission accomplishes spontaneous modulation onto the electron beam and induces an energy modulation amplitude up to a factor of 8.5. Energy modulation amplitude is defined as the laser-induced modulation divided by the slice energy spread. Then, the amplified emission is directed to the laser transmission line. At the same time, the electrons pass the second chicane. Appropriate time delays are introduced here so that the electron beam and emission are again precisely time-synchronized at the entrance of the next modulator. The laser transmission line is an optical system consisting of mirrors and lenses. It is used to transmit and focus the optical field while minimizing power loss. The power loss is assumed to be 5\% in the simulations. In reality, more significant power loss may occur. If that happens, then the third-stage modulation amplitude is reduced, and the dispersion strength needs to be dynamically adjusted to optimize the FEL radiation. In the third modulator, the reused seed laser produces an energy modulation amplitude of up to 6. The last two stages of modulation of the e-beam are concluded by its coherent emission, implying excellent synchronization. Figure \ref{fig5} shows the profiles of laser-induced modulation amplitude in the second and third modulators.

\begin{figure*}[!htb]
\begin{indented}
\item[]
    \includegraphics[width=\linewidth]{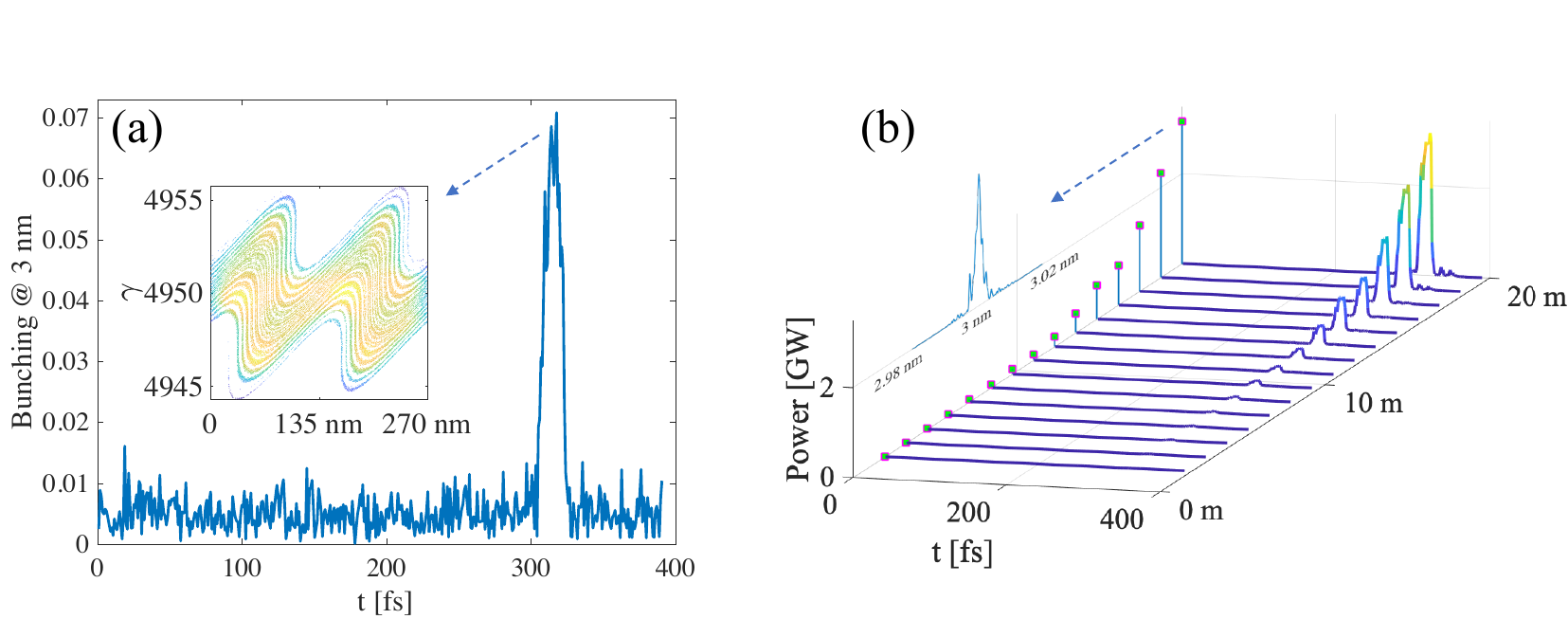}
    \caption{Performances of HOK-EEHG at 3 nm: (a) bunching factor along the beam and phase space at the entrance of radiator; (b) evolution of the radiation power along the radiator and spectrum at 20 m.}
    \label{fig6}
    \end{indented}
\end{figure*}

To optimize bunching at 3 nm, the n=-2 tune is adopted. The dispersion strengths of the last two chicanes are set as 1.37 mm and 0.05 mm, respectively. At the exit of the third chicane, the distribution of the bunching factor and longitudinal phase space of the electron beam are provided in Fig.~\ref{fig6}(a). The maximum bunching factor along the electron beam is 7.1\%, and there is a well-defined phase space structure at the maximum. The separated energy bands look very similar to those of a standard EEHG, demonstrating the feasibility of the proposed approach. The radiation power grows exponentially, reaching saturation at about 20 m. At saturation, the FEL radiation has a peak power of 3.2 GW, a pulse length of 14.6 fs (FWHM), and a spectral bandwidth ($\Delta\lambda/\lambda$) of about 1/2068. The bandwidth is 1.6 times the Fourier transform limit, indicating great temporal coherence. SASE radiation, which is also produced during the seeding amplification, is less than 0.1 MW at 20 m and barely has no apparent influence on the spectrum. These results are shown in Fig.~\ref{fig6}(b).
The above simulation results show that the proposed approach can generate nearly fully coherent soft X-ray radiation with a wavelength of 3 nm, a peak power of several gigawatts, and a pulse duration of about fifteen femtoseconds.

\section{Conclusions and outlook}

Seeded FELs are compelling tools to generate almost fully coherent radiation at EUV and soft X-ray range. The proposed technique combines the advantages of HOK and EEHG, thus enabling the generation of ultra-high harmonic radiation. Simulations give strong evidence of the proposed approach for producing coherent X-ray radiation at 3 nm. They can further extend the photon wavelength coverage to 1.5 nm, mainly by increasing the harmonic number of HOK. With radiation down to 3 nm or even shorter, seeded FEL with laser-like attributes will serve as an integrated platform for the cutting-edge ultrasmall sciences.

Seeded FELs, constrained by the power of seed laser, are typically unable to operate at the same high repetition rate as the e-beam from the superconducting linac. In nominal HGHG or EEHG, the peak power of the seed laser is usually on the order of hundreds of megawatts. In the proposed approach, however, the peak power of the external seed is 1.78 MW. Lower peak power can be exchanged for a higher repetition rate. It is expected that even with 1 MHz repetition rate, the average power of the seed laser is only 0.62 W with a pulse length of 350 fs (FWHM). These parameters could be satisfied by a commercial Ti:sapphire laser or Yb-based fiber laser system, presenting a great opportunity to achieve high-repetition-rate seeded FELs.

\section{Acknowledgements}

This work is supported by the Shenzhen Science and Technology Program (Grant No. RCBS20210609104332002), the Scientific Instrument Developing Project of Chinese Academy of Sciences (Grant No. GJJSTD20220001), and the National Natural Science Foundation of China (Grant No. 22288201).


\section*{References}
\bibliographystyle{unsrt}
\bibliography{mybib}

\end{document}